\pdfoutput=1
\documentclass[12pt,leqno]{article}
\usepackage{fullpage}
\usepackage{fancyhdr}
\usepackage{amsfonts}
\usepackage{amsmath}
\usepackage{mathrsfs}

\title{Hyperbolic Multi-Monopoles \\ With Arbitrary Mass}
\author{Lesley M. Sibner\footnote{Department of Mathematics, Polytechnic Institute of NYU, Brooklyn, NY 11201, USA \newline \hspace*{0.25 in} lsibner@duke.poly.edu } \hspace*{1 pt} and Robert J. Sibner\footnote{Department of Mathematics, City University of New York, Graduate Center and \hspace*{0.25 in} Brooklyn College,  Brooklyn, NY 11210, USA.  rsibner@gc.cuny.edu} \footnote{Research partially supported by PSC-CUNY grant 69570-00-38}}
\date{October 1, 2012}
\begin{document}

\addtocounter{page}{-1}
\maketitle
\begin{abstract}
 On a complete manifold, such as $\mathbb{R}^{3}$ or hyperbolic space $\mathbb{H}^{3}$, the limit at infinity  of the norm of the Higgs field $\Phi$ is called the \textit{mass} of the monopole.  We show the existence, on $\mathbb{H}^{3}$, of monopoles with given magnetic charge and arbitrary mass.  Previously, aside from charge one monopoles, existence was known only for monopoles with integral mass (since these arise from U(1) invariant instantons on $\mathbb{R}^{4}$). The method of proof is based on Taubes' gluing procedure, using well-separated, explicit, charge one monopoles.  The analysis is carried out in a weighted Sobolev space and necessitates eliminating the possibility of point spectra.
\end{abstract}
\thispagestyle{empty}
\newpage
\section{Higgs Field and Holonomy}

The characteristic numbers of a magnetic monopole on a 3-manifold M (without
 boundary) are its integer valued magnetic charge k and the asymptotic limit m
of the length of its Higgs field at an end.  (This limit is referred 
to as the \textit{mass} of the Higgs field.)  For M $= \mathbb{R}^{3}$, using 
scaling techniques, one can assume without loss of generality that $m = 1$, but 
on hyperbolic space $\mathbb{H}^{3}$ scaling is not possible.  Aside from known explicit 
\textit{charge one} monopoles, most attention has been given to monopoles on $\mathbb{H}^{3}$ with 
integral m.  The reason for this is that one can (see Atiyah [A]), up to conformal
equivalence, consider $\mathbb{H}^{3}$ as $\mathbb{R}^{4}\setminus\mathbb{R}^{2}$
modulo a U(1) action leaving $\mathbb{R}^{2}$ invariant.  Then any U(1) invariant 
instanton on $\mathbb{R}^{4}$  
with fixed set $\mathbb{R}^{2}$ produces a \textit{hyperbolic monopole} on $\mathbb{H}^{3}$
with integral m.  Conversely, a hyperbolic monopole on $\mathbb{H}^{3}$ can be 
lifted to an instanton on $\mathbb{H}^{3} \times$ S$^{1}$, resulting in a 
\textit{hyperbolic caloron} (see [A, L, H, MS]).  We have shown previously that it is precisely
the hyperbolic monopoles with integral mass m which allow extensions to 
instantons on all $\mathbb{R}^{4}$.  To understand this better, we recall [SS1, SS2] that 
any L$^{2}$ (singular) connection on $\mathbb{R}^{4}\setminus\mathbb{R}^{2}$ (not 
necessarily satisfying any field equations) has a well-defined limit holonomy which
is constant along the singular set $\mathbb{R}^{2}$.  (The L$^{2}$ connections
are classified by the space of flat connections.)  The limit  holonomy corresponds 
(via Atiyah's construction) to the mass of the corresponding configuration on 
$\mathbb{H}^{3}$, with integral mass corresponding to integral holonomy.  The integrality
of the holonomy is a necessary and sufficient condition for the connection on 
$\mathbb{R}^{4}\setminus\mathbb{R}^{2}$ to extend across the $\mathbb{R}^{2}$.  
On the other hand, however, from the point of view of the 3-manifold $\mathbb{H}^{3}$,
the restriction to integral m seems clearly artificial.  Indeed, on $\mathbb{H}^{3}$
it makes analytic sense to prescribe the mass as any positive real number.  This view was already pointed out by Atiyah in his 1984 publication [A], and is supported by the example [FHP1, FHP2] of a U(1) invariant instanton on 
$\mathbb{R}^{4}\setminus\mathbb{R}^{2}$ with holonomy $1/2$ which does \textit{not} 
extend.  \\

We recall some basic definitions.  A connection on an SU(2) vector bundle over
a simply connected 3-manifold M can be pulled back to an su(2) valued (connection)
1-form A on M and gives rise to a covariant derivative $d_{A} = d + [A, \;]$.  A 
\textit{Higgs monopole} on M (see [JT]) is a configuration pair $c = (\Phi,A)$ where $\Phi$ is an su(2)
valued function on M.  (If M is not simply connected, these pullbacks are local.) 
The curvature of the connection is given by $F = dA + \frac{1}{2}[A,A]$ and the pair
satisfies the Bogomolny monopole equations [Bo]:
\begin{equation}\tag{1.1}
 d_{A}\Phi = *F_{A}
\end{equation}
The solutions of (1.1) are the absolute minima of the Yang-Mills-Higgs action 
functional
\begin{equation}\tag{1.2}
 \mathcal{YMH}(c) = \frac{1}{2}\int\limits_{M}(|F_{A}|^{2}+|\;d_{A}\Phi|^{2})\;dV
\end{equation}
This theory becomes interesting if M has an ``end'' (such as $\mathbb{R}^{3}$ 
or $\mathbb{H}^{3}$) in which case the natural boundary and topological conditions are given 
by prescribing the mass of the monopole
\begin{equation}\tag{1.3}
 m = \lim\limits_{|x|\rightarrow \infty}|\Phi(x)|
\end{equation}
and the magnetic charge
\begin{equation}\tag{1.4}
 k = \dfrac{1}{4\pi m}\int\limits_{M}tr(F_{A} \wedge \; d_{A} \Phi)
\end{equation}
Our main result is the following existence theorem for multi-monopoles on
$\mathbb{H}^{3}$ of arbitrary mass.

\begin{flushleft}
 \textbf{\underline{Theorem.}} There exists a smooth configuration $c = (\Phi,A)$ on $\mathbb{H}^{3}$,
having prescribed magnetic charge $k \in \mathbb{Z}$, prescribed mass $m > 0$ at 
infinity, and satisfying the Bogomolny monopole equation (1.1).
\end{flushleft}

   From the preceding remarks, noting the identification of mass for monopoles on $\mathbb{H}^{3}$ with holonomy for instantons on $\mathbb{H}^{3} \times S^{1}$, it is clear that one has immediately the following:\\
   
\noindent  \textbf{\underline{Corollary.} }   There exists a hyperbolic caloron (on M $= \mathbb{H}^{3} \times  S^{1}$) having prescribed topological charge (instanton number) N and prescribed holonomy (not necessarily integral).\\

In the mid 1970's, by making a symmetry ansatz, an explicit SU(2) Bogomolny 
monopole solution on $\mathbb{R}^{3}$ was obtained by Prasad and Sommerfeld [PS].
(Earlier, Dirac [Di] had obtained a U(1) monopole on $\mathbb{R}^{3}$.)  Later,
Chakrabarti [C] wrote down an explicit solution on hyperbolic 3-space 
$\mathbb{H}^{3}$ (see also Braam [Br]).  All of these explicit solutions were of 
charge $k = 1$.  The explicit solutions on $\mathbb{R}^{3}$ became ``building blocks''
in the next development.  With a procedure that has, by now, become standard, Taubes 
[JT] constructed monopoles on $\mathbb{R}^{3}$ (with arbitrary charge) by a 
``patching argument.''  The basic idea (which we will follow in the sequel) is
to construct an ``approximate'' monopole and then perturb it to a true monopole
using some version of the Implicit Function Theorem.  This method gives solutions
which are (for given charge) absolute minima of the action functional.  (Taubes went
on to utilize this program to show that the moduli space of self dual instantons
on appropriate 4-manifolds was non empty.)  The program has been used by Floer [F1, F2],
Ernst [E] and Durenard [Du] in the construction of monopoles on 3-manifolds with
Euclidean ends.  These last results, together with our Theorem, naturally suggest (with some restrictions) the following:\\

\noindent \textbf{\underline{Conjecture.}}  Let $M$ be a Riemannian 3-manifold with $N$ hyperbolic ends.  There exists 
a smooth configuration on $M$, satisfying the Bogomolny equations, having prescribed
magnetic charge $k \in \mathbb{Z}$ and prescribed asymptotic limits $m_{j}$, 
$j = 1$, ..., $N$ at the ends. \\

To prove the theorem, we follow a program similar 
to that developed by Taubes and discussed above.  We work directly
on the space $\mathbb{H}^{3}$ and do not require any assumption on the magnitude
of $m$ nor on the charge $k$. \\

In Section 2 we construct an approximate solution and define the weighted Sobolev space
in which it lies.  In Section 3 the perturbation problem is derived.  Sections 
4 and 5 are devoted to showing that the origin is not contained in the spectrum 
of the various operators.  In section 4 we use results of Mazzeo [M] to show the absence of eigenvalues.  In section 5 we show that the essential spectra have positive lower bounds.  Then, in Section 6, a lower bound estimate is obtained
for a linearized operator.  A continuity argument completes the proof of existence.\\

In light of recent interest in the classification of hyperbolic monopoles by the asymptotic values of their Higgs fields, it is perhaps worth noting that all the monopoles that we construct are of Dirac type at infinity.\\

We remark that this general method is, by now, standard and has been exploited 
successfully in many situations in which the lower bound can be established.  
However, in our case, the a priori bound is false in the usual L$^{2}$
Sobolev spaces over $\mathbb{H}^{3}$ and, in order to work directly on $
\mathbb{H}^{3}$, one must use weighted Sobolev spaces.\\

The authors wish to acknowledge and thank the graciousness of the Max Planck Institute in Bonn, Germany and the Institut des Hautes Etudes Scientifique in Bures-sur-Yvette, France for their hospitality during the periods when preliminary work on this project was carried out. 

\section{The Approximate Monopole}
The basic SU(2), charge $k=1$ and mass m, monopole on $\mathbb{H}_{3}$ can be written
 down explicitly [C].  To obtain an approximate monopole of given charge, $k$, first 
choose $k$ points $x_{1}, ..., x_{k}$ as centers of charge one monopoles, requiring that
the distance between centers be greater than 6$R$ where $R \geq 1$ is a constant to 
be chosen later.  The distance between the centers $x_{i}$ are also chosen so that, 
denoting by $B_{\rho}^{\text{i}}$ the ball of radius $\rho$ about $x_{i}$, we can 
choose geodesic spherical coordinates $(r_{i}, \theta_{i}, \chi_{i})$ centered at 
$x_{i}$ so that the half rays $\theta_{i} = 0$ and $\theta_{i} = \pi$ do not
intersect the closure of any of the sets $U_{R}^{j} = B_{2R}^{j} \setminus B_{R}^{j}$ 
for $j \neq i$.\\

Let $c_{i} = (\Phi_{i}, A_{i})$ be the basic Chakrabarti, charge one, monopole at $x_{i} 
\in \mathbb{H}^{3}$ [C].

\begin{equation}\tag{2.1a}
 \begin{cases}
  \Phi_{i} = (\alpha \coth \hspace{1 pt}\alpha r_{i} - \coth \hspace*{-0.02 in}r_{i})\hat{i}\\
  A_{i} = \dfrac{\alpha \sinh \hspace*{0.01 in} r_{i}}{\sinh\alpha r_{i}}(\; d\theta_{i}\hat{j} + 
    \sin \theta_{i} \; d\begin{LARGE}\chi_{i}\end{LARGE}\hat{k}) + (1 - \cos \theta_{i}) \; d \chi_{i}\hat{i}
 \end{cases}
\end{equation}
where we have written $\alpha = m + 1$ and the Pauli matrices $\hat{i}, 
\hat{j}, \hat{k}$ are a basis for su(2).  \\

In the neighborhood of infinity $N_{\infty} = \mathbb{H}^{3} \setminus 
\cup_{i=1}^{k} B_{R}^{i}$, we take a U(1)-Dirac monopole $c_{\infty} = 
(\Phi_{\infty},A_{\infty})$ where\\
\begin{equation}\tag{2.1b}
\begin{split}
\Phi_{\infty} &= \lbrace (\alpha - 1) + (1 - \coth\hspace*{-0.02 in} r_{1}) + ... + (1 - \coth\hspace*{-0.02 in} r_{k}) \rbrace \hat{i} \\
A_{\infty} &= \lbrace (1 - \cos\theta_{1})d\begin{LARGE}\chi_{1}\end{LARGE} + ... + (1 - \cos \theta_{k})d\begin{LARGE}\chi_{k}\end{LARGE}) \rbrace \hat{i}.
\end{split}
\end{equation}

\indent In any system of geodesic polar coordinates $(r, \theta, \chi)$ the metric is
given by 
\begin{align*}
 &ds^{2} = dr^{2} + \sinh^{2}\hspace*{-0.04 in}rd\Omega^{2}&\\
 &\hspace*{0.2in} = dr^{2} + \sinh^{2}\hspace*{-0.04 in}r \; d\theta^{2} + \sinh^{2}\hspace*{-0.04 in}r\sin^{2}\hspace*{-0.04 in}\theta \; d\chi^{2}&
\end{align*}
so that the volume element is given by $dV = \sinh^{2}r\sin\theta \; dr \; d\theta \; d\chi$.
 In the metric induced on the cotangent space
\begin{equation}\tag{2.2}
 |\; dr| = 1, |\; d\theta| = (\sinh\hspace*{-0.02 in} r)^{-1} \text{ and } |\; d\chi| = (\sinh \hspace*{-0.02 in}r \sin\theta)^{-1}
\end{equation}
\indent While the Higgs action of $c_{i}$ $(1 \leq u \leq k)$ is finite, the action 
of $c_{\infty}$ on $\mathbb{H}^{3}$ is \textit{not} finite because of the singular
behaviour of $F_{A_{\infty}}$ at the points $x_{i}$.  However the restriction of 
$F_{A_{\infty}}$ to $N_{\infty}$ does have finite action.  In the gauge of (2.1a) 
$c_{i}$ has a Dirac string singularity along the half ray $\theta_{i} = \pi$.
(note that $|1-\cos\theta_{i}||\; d\chi_{i}| \rightarrow \infty$ as $\theta_{i} \rightarrow \pi$).
However, since the holonomy around the string is integral, the codimension two 
removable singularity theorem [SS1,SS2] ensures that $c_{i}$ is gauge equivalent to a 
smooth configuration.  The same is true of $c_{\infty}$ in $N_{\infty}$.\\

We emphasize that $c_{\infty}$ and each of the $c_{i}$ are solutions of the 
Bogomolny equations: 
\begin{equation}\tag{2.3}
 d_{A}\Phi = *F_{A}.
\end{equation}

In particular, this is true for $c_{i}$ on $B_{2R}^{i}$ and $c_{\infty}$
on $N_{\infty}$, so that if we glue them together by a partition of unity
$\lbrace \lambda_{1}, ..., \lambda_{k}, \lambda_{\infty} \rbrace$ subordinate
to the covering of $\mathbb{H}^{3}$ by $N_{\infty}$ and the $B_{2R}^{i}$, 
$1 \leq i \leq k$, we obtain an ``approximate monopole'' 
$c_{0} = (\Phi_{0}, A_{0}):$
\begin{equation}\tag{2.4}
 \begin{cases}
  \Phi_{0} = \lambda_{\infty}\Phi_{\infty} + \sum\limits_{i=1}^{k}\lambda_{i}\Phi_{i}\\
  A_{0} = \lambda_{\infty}A_{\infty} + \sum\limits_{i=1}^{k}\lambda_{i}A_{i}
 \end{cases}
\end{equation}

\indent By its construction, $c_{0}$ satisfies the monopole equation (2.3) in 
each $B_{R}^{i}$ and in $N_{\infty} \setminus \cup_{i=1}^{k}B_{2R}^{i}$.
We need to estimate the deviation of $c_{0}$ from a solution in the intersections
$U_{R}^{i}= B_{2R}^{i} \setminus B_{R}^{i}$.  Note that, in the partition of 
unity construction, at most two $\lambda$'s can be non-zero simultaneously.  
In particular, in $U_{R}^{i}$ we have $\lambda_{i} + \lambda_{\infty} = 1$. 
Moreover, for $x \in U_{R}^{i}$ and $k \neq i$, one has $r_{k}(x) > 4R$.
In $U_{R}^{i}$\vspace*{0.05 in}
\begin{flushleft}
 $(2.5)$  $d_{A_{0}}\Phi_{0} - *F_{A_{0}} = \lambda_{\infty}(1-\lambda_{\infty})
\lbrace [A_{\infty}-A_{i},\Phi_{\infty}-\Phi_{i}] + *[A_{\infty}-A_{i},A_{\infty}-A_{i}] \rbrace $\\
\hspace*{1.2 in}$+ \; d\lambda_{\infty}(\Phi_{\infty}-\Phi_{i})-*(\; d\lambda_{\infty} \wedge (A_{\infty}
-A_{i}))$
\end{flushleft}
\vspace*{0.05 in}
\indent The terms in (2.5) can be estimated, using (2.2), to obtain the pointwise
bound in $U_{R}^{i}$:

\begin{equation}\tag{2.6}
 |\; d_{A_{0}}\Phi_{0} - *F_{A_{0}}| \leq K (e^{-\alpha r_{i}} + e^{-4R})
 \text{ with } K \text{ independent of } R \geq 1.
\end{equation}

\indent  This can be done in the subdomain of $U_{R}^{i}$ where $0 \leq |\theta_{i}| 
< 3\pi / 4$ in the gauge in which the configuration is represented by (2.1ab).
In the overlapping region where $\pi / 4 < |\theta_{i}| \leq \pi$ one should
choose a gauge in which the string is given by $\theta_{i} = 0$.  See [JT] for 
a discussion of inverting strings and also for the fact that the existence of local 
smoothing gauges implies the existence of a global smoothing gauge.  (The estimate
(2.6) is gauge invariant.)  In computing bounds for each term of (2.5) in the region
$U_{R}^{i}$, one finds that it is the term $|A_{\infty} - A_{i}|$ that decays most
slowly and gives the upper bound (2.6).\\

\indent We define the weighted spaces $L_{\beta}^{p}$ on $q$-forms, as the completion
of $C_{0}^{\infty}(\wedge^{q})$ in the norm:
\begin{center}
 $||\omega||_{p,\beta} = (\int\limits_{\mathbb{H}^{3}}|\omega|^{p}\cosh^{2}\hspace*{-0.04 in}\beta r \; dV)^{1/p}
$
\end{center}
We assume $\beta < 1$ which ensures that our appproximate monopole $c_{0}$ has finite 
weighted action; namely,
\begin{center}
 $\mathcal{YMH}_{\beta}(c_{0}) = \dfrac{1}{2}\int\limits_{\mathbb{H}^{3}}(|F_{A_{0}}|^{2} + 
|\; d_{A_{0}}\Phi_{0}|^{2})\cosh^{2}\hspace*{-0.04 in}\beta r\; dV < \infty$.
\end{center}

We can now easily show\\

\noindent \textbf{Proposition 2.7.} If $\beta <$ min$(1,m)$, there is a constant
$c > 0$ depending on $m$ but not on $R$, such that
\begin{center}
$||\; d_{A_{0}}\Phi_{0} - *F_{A_{0}} ||_{2,\beta} \leq Ke^{-cR}$
\end{center}
\textit{Proof}:  Outside $\cup_{i=1}^{k}U_{R}^{i}$, $d_{A_{0}}\Phi_{0} - *F_{A_{0}} $
 vanishes.  Using (2.6) in $U_{R}^{i}$ and recalling that $\alpha = m + 1$, we see
that 
\begin{center}
$||\; d_{A_{0}}\Phi_{0} - *F_{A_{0}} ||_{2,\beta}^{2} \leq 
K$ \begin{LARGE}$\lbrace$\end{LARGE} $\int\limits_{R}^{2R}e^{(-2m+2\beta)r_{i}}\; dr_{i}
 + e^{-8R}(\int\limits_{R}^{2R}e^{2\beta+2)r_{i}}\; dr_{i})$ \begin{LARGE}$\rbrace$\end{LARGE}\\
 $\leq K$ \begin{LARGE}$\lbrace$\end{LARGE} $e^{4(\beta-m)R} + e^{4(\beta-1)R}$ \begin{LARGE}$\rbrace$\end{LARGE}
\end{center}
The constant $c$ will have the right sign if $\beta < m$ and $\beta < 1$.  This proves
the proposition.

\section{The Perturbation Problem}
We now look for solutions of (1.1) of the form $c = c_{0} + \zeta = 
(\Phi_{0} + \varphi, A_{0} + a$) with $c_{0}$ the approximate monopole of $\S$2.
Following Floer [F2], we expand (1.1) in a Taylor expansion to obtain
\begin{equation}\tag{3.1}
 Lc = \; d_{A}\Phi - *F_{A} = Lc_{0} + D_{L}\zeta + \sigma(\zeta,\zeta)
\end{equation}
\indent Here, $D_{L}$ is the linearization of $L$ at $c_{0}$; i.e.,
\begin{equation}\tag{3.2}
D_{L}\zeta = D_{L}(\varphi,a) = - \;*d_{A_{0}}a + \;d_{A_{0}}\varphi - [\Phi_{0},a] 
\end{equation}
The quadratic term $\sigma(\zeta,\zeta)$ is defined by a bilinear bundle map 
of $\wedge^{0} \oplus \wedge^{1}$ into $\wedge^{1}$.\\
\indent We work in the weighted Sobolev spaces described in $\S$2.  Throughout, the 
mass is a fixed arbitrary positive number and the constant $\beta$ appearing in the
weight factor satisfies $0 < \beta < \text{min}(1,m)$.  In the weighted Sobolev space,
the adjoint of the operator $d_{A_{0}}$ on forms $\omega \in \wedge^{q}$ is:
\begin{center}
 $d_{A_{0}}^{\dagger} \omega = (\cosh^{-2}\hspace*{-0.04 in}\beta r)\;d_{A_{0}}^{*}  ((\cosh^{2}\hspace*{-0.04 in}\beta r)\omega)$
\end{center}
where $d_{A_{0}}^{*}$ is the ordinary $L^{2}$ adjoint.
To obtain ellipticity, we add a ``slice'' condition:
\begin{equation}\tag{3.3}
 D_{S}\zeta = D_{S}(\varphi,a) = d_{A_{0}}^{\dagger}a - [\Phi_{0},\varphi]=0
\end{equation}
This together with (3.2) gives an elliptic operator on pairs
\begin{center}
$ \delta = (D_{S},D_{L}) : \wedge^{0} \oplus \wedge^{1} 
\rightarrow \wedge^{0} \oplus \wedge^{1}$
\end{center}
defined by 
\begin{equation}\tag{3.4}
 \delta \zeta = (\; \delta_{A_{0}} - ad\Phi_{0})(\varphi,a)
\end{equation}
where we have written $\delta_{A_{0}}(\varphi,a) = (\; d_{A_{0}}^{\dagger}a, \;d_{A_{0}}\varphi-
*\;d_{A_{0}}a)$ and $ad\Phi_{0}(\varphi,a) = ([\Phi_{0},\varphi],[\Phi_{0},a])$.\\

\noindent To solve the Bogomolny equations (1.1), we want to find solutions 
$\zeta = (\varphi,a)$ of

\begin{equation}\tag{3.5}
 \delta\zeta + \zeta \# \zeta = G_{0}
\end{equation}

\noindent where $\zeta \# \zeta = (0,\sigma(\zeta,\zeta))$ and $G_{0} = (0,-Lc_{0})$ 
is sufficiently small in appropriate norms.

\noindent As in [T, FU], we look for a solution perpendicular to the kernel of 
$\delta$ by setting $\zeta = \delta^{\dagger}\eta$ and solving for 
$\eta = (\psi,b), \psi \in \wedge^{0}$ and $b \in \wedge^{1}$,
\begin{equation}\tag{3.6}
 \delta\delta^{\dagger}\eta + \delta^{\dagger}\eta\#\delta^{\dagger}\eta = G_{0}.
\end{equation}
 
\noindent Since $ad\Phi_{0}$ is skew adjoint,
\begin{equation}\tag{3.7}
\delta^{\dagger} = \delta_{A_{0}}^{\dagger} + ad\Phi_{0}
\end{equation}
\noindent so that
\begin{equation}\tag{3.8}
\delta\delta^{\dagger} = (\delta_{A_{0}} - ad\Phi_{0})(\delta_{A_{0}}^{\dagger}+ad\Phi_{0}) = \delta_{A_{0}}\delta_{A_{0}}^{\dagger} - (ad\Phi_{0})^{2} + E
\end{equation}

\noindent where $E\eta = (- * [d_{A_{0}}\Phi_{0},*b],[d_{A_{0}}\Phi_{0},\psi] - 
*[d_{A_{0}}\Phi_{0},b])$.\\

\noindent A computation shows that the $L_{\beta}^{2}$ adjoint of $\delta_{A_{0}}$ is:
\begin{equation}\tag{3.9}
 \delta_{A_{0}}^{\dagger}\eta = \delta_{A_{0}}^{\dagger}(\psi,b) = (d^{\dagger}_{A_{0}}
b,d_{A_{0}}\psi-*(\cosh^{-2}\hspace*{-0.04 in}\beta r) b) d_{A_{0}} (\cosh^{2}\hspace*{-0.04 in}\beta r)b
\end{equation}

\noindent It follows that
\begin{equation}\tag{3.10}
 \delta_{A_{0}}\; \delta_{A_{0}}^{\dagger}(\eta)=d_{A_{0}}d_{A_{0}}^{\dagger}+d^{\dagger}_{A_{0}}d_{A_{0}} + T_{A_{0}}\eta = \Delta_{A_{0}}\eta + T_{A_{0}}\eta 
\end{equation}

\noindent where,  $\Delta_{A_{0}}\eta = \Delta_{A_{0}}(\psi,\text{b}) = (\Delta_{A_{0}}\psi, \Delta_{A_{0}}\text{b})$ and 
\begin{equation}\tag{3.11}
 T_{A_{0}}\eta = (*[F_{A_{0}},b],*[F_{A_{0}},\psi]-2\beta\tanh \beta r\lbrace* (dr \wedge * \; d_{A_{0}}b)
+\;d^{*}_{A_{0}}(b \wedge  \; dr)\rbrace).
\end{equation}

When $A_{0} = 0$, these operators reduce, respectively, to the (weighted) scalar Laplacian $\Delta$ and to 
\begin{equation}\tag{3.12}
T= (0,T_{1}) \text{ where } T_{1} = -2\beta \tanh \; \beta r \lbrace \ast dr\wedge\ast db + d^{\ast}(dr \wedge b)\rbrace
\end{equation}

\noindent Writing
\begin{equation}\tag{3.13}
 \Delta_{A_{0}} = \Delta + S \text{ and } T_{A_{0}} = T + R 
\end{equation}
\noindent one finds \\
\vspace*{-0.25 in}
\begin{equation}\tag{3.14}
 S\eta = (S_{0}\psi, S_{1} b) \text{ where } S_{0}\psi = d^{\dagger}[A_{0},\psi]+\*[A_{0},d^{\dagger}b+\*[A_{0},\*b]] \text{ and}
\end{equation}

\hspace*{-0.25 in} $S_{1}b = d^{\dagger}[A_{0},b] + d(*[A_{0},*b])+*[A_{0},*(db + [A_{0},b])] + [A_{0},d^{\dagger}b+*[A_{0},*b]]$\vspace*{-0.25 in}

\begin{equation}\tag{3.15}
 R\eta = (*[F_{A_{0}},b],*[F_{A_{0}},\psi] + * 2\beta  \tanh \; \beta r \lbrace[A_{0},*(dr \wedge b)]
- dr \wedge *[A_{0},b]\rbrace)
\end{equation}

Also, for $A_{0} = 0$ we see from (3.10) that
\begin{equation}\tag{3.16}
 \delta_{0}\delta_{0}^{\dagger} = \Delta + T
\end{equation}

Moreover, using (3.10), (3.13) and (3.16),
\begin{equation}\tag{3.17}
 \delta_{A_{0}}\delta_{A_{0}}^{\dagger} = \Delta_{A_{0}} + T_{A_{0}} = \Delta + R + S + T = \delta_{0}\delta_{0}^{\dagger} + R + S
\end{equation}

Recall from (2.1b) that, near infinity, $c_{0} = (\Phi_{0},A_{0}) = (\Phi_{\infty}, A_{\infty})$ so that this configuration decays exponentially in the sense that 
\begin{center}
 $|d\Phi_{0}|, |A_{0}|, |F_{A_{0}}| \leq e^{-c|x|}$ for large $|x|$
\end{center}

\noindent As a result, an examination of the various ``remainder'' operators above shows that they are all rapidly decaying.\\

\noindent \textbf{\underline{Proposition 3.18.}} $|(R + S + T + E)\eta| \leq \tau(r)(|\eta| + |\nabla \eta|)$\\ with $\tau(r)$ decaying exponentially at infinity.

\section{The Absence of Point Spectra}

\noindent We will see in Section 5 that the weight factor, $\cosh^{2}\beta r$, 
shifts the essential spectrum of the Laplacian to the right.  The same is true for
the operator $\delta\delta^{\dagger}$ which differs from the Laplacian by a first 
order partial differential operator.  To resolve the question of the possible 
emergence of eigenvalues when the weight is introduced, we note that the weighted norm
comes from the spherically symmetric metric
\begin{center}
 $ds^{2} = dr^{2} + f^{2}(r) \; d\Omega^{2}$
\end{center}
where $f(r) = \cosh\beta r \sinh r$.  The non-zero components of the Riemann
curvature tensor are:
\begin{center}
 $\dfrac{-f''}{f} \text{ and } \dfrac{1-f'^{2}}{f^{2}}$.
\end{center}

Since $f(0) = f''(0) = 0$ and $f'(0) = 1$, the metric is non-singular, and the 
sectional curvatures are bounded.  Therefore, in general (with $d_{B}^{\ast}$ the adjoint of $d_{B}$).\\
\indent $(i)$ \hspace*{0.25 in} $d_{B_{i}} - \nabla_{i} = B_{i}$\\
\indent $(ii)$ \hspace*{.2 in} $\Delta_{B} = d_{B} \; d_{B}^{*}+
\; d_{B}^{*} \; d_{B} = \nabla^{*}\nabla + Q$\\

\noindent where $Q$ involves $F_{B}$ and Ricci curvature and satisfies:
$|Qu| \leq c(|u| + |\nabla u|)$.  (See [JT] for more discussion.)  \\

Let $M = (\mathbb{H}^{3},g)$ where $g$ is the metric above.  Let $E$ be
a vector bundle over $M$ with structure group SU(2).  In the following sections, E will be either the bundle of zero-forms or of one-forms.\\

\noindent \textbf{\underline{Theorem 4.1.}} (Mazzeo [M]) Let $u \in C^{\infty}(M, E).$
Suppose $u=0$ when $r \leq r_{0}$, $|u| = O(e^{-cr})$ for some $c >0$, and $u$ is in the 
$L^{2}$ domain of $\nabla^{*}\nabla$.  Then, for $r_{0}$ and $t$ sufficiently large and 
$C$ independent of $t$,

\begin{center}
 $t^{3}\int e^{2tr}|u|^{2}\;dV + t \int e^{2tr}|\nabla u|^{2}\;dV \leq 
C \int e^{2tr}|\nabla^{*}\nabla u|^{2}\;dV$.\\
\end{center}

\noindent \textbf{\underline{Corollary 4.2.}} Let $u$ be as in Theorem 4.1.  If, in addition 
\begin{center}
 $|\nabla^{*}\nabla u| \leq k (|u| + |\nabla u|)$
\end{center}
then $u \equiv 0$.\\

\noindent \textbf{\underline{Corollary 4.3.}} Let $w \in L_{\beta}^{2}(\mathbb{H}^{3},E)$ 
be an eigenfunction satisfying 
\begin{center}
$Lw = \lambda w$\\
\end{center}

\noindent where $L = \nabla^{*}\nabla + Q$ where $Q$ satisfies the bound in $(ii)$ above.  Then $w \equiv 0$.\\

\indent By Proposition (3.18) we know that all of the operators we will consider in the following sections (namely $\Delta,
 \delta_{0}\delta_{0}^{\dagger}, \delta_{A_{0}}\delta_{A_{0}}^{\dagger}$, or 
$\delta\delta^{\dagger})$ satisfy the hypotheses of Corollary 4.3 and, consequently, have no point spectrum.    In all cases, the spectrum consists
only of essential spectrum.  Nevertheless, for consistency with quoted results and for clarity of exposition, although they coincide in our case, in the following we will use $\sigma$ to denote the spectrum of an operator and $\sigma_{\infty}$ its essential spectrum.

\section{The Essential Spectrum}
To calculate the essential spectrum, we apply Donnelly's method [Do] of separation
of variables to the weighted space and to sections (zero forms or one forms) with 
su(2) valued coefficients.\\

Recall that $\Delta_{p} = d^{\dagger}d + dd^{\dagger}$ ($p = 0$ or 1) is the weighted self-adjoint 
Laplacian on $L_{\beta}^{2}(\mathbb{H}^{3})$ with domain $L_{\beta}^{2,2}(\mathbb{H}^{3})$ and $\delta_{0}\delta_{0}^{\dagger}$ is the Floer operator of (3.6) (with $A_{0} = 0)$.\\

\noindent \textbf{\underline{Proposition 5.1.}}\\
\indent$(a)$ $\sigma(\Delta_{0}) = [(1+\beta)^{2}, \infty)$\\
\indent$(b)$ $\sigma(\Delta_{1}) = [\beta^{2},\infty)$  \\
\indent$(c)$ $\sigma(\delta_{0}\delta_{0}^{\dagger}) = [\beta^{2},\infty)$\\

\noindent We emphasize that this result applies to the operators evaluated at the 
zero connection.\\

We make extensive use of the following proposition which tells us that, as long
as the coefficients tend to zero at infinity, a smooth first order operator $C$ is 
relatively compact with respect to a self-adjoint second order elliptic operator $L$; its 
addition to $L$ does not change the essential spectrum.  We state this in the form 
which will be applied.\\

\noindent \textbf{\underline{Proposition 5.2.}} Let $L$ be an elliptic, second order,
self-adjoint operator on $L_{\beta}^{2}(M)$, where $M$ is either the non-negative reals 
$\mathbb{R}^{+}$, or $\mathbb{H}^{3}$.  Assume that the domain of $L$ is $\mathcal{H}_{B} =
L_{\beta, B}^{2,2}$ (using covariant derivatives at B).  Let $C = \sum a_{i}\frac{\partial}{\partial x_{i}}+ b$ where the coefficients
are smooth functions and $\tau(r) =$ max$_{|x| \geq r}(|a_{i}|,|b|)$.  If $\tau(r)$ 
tends to zero as $r$ tends to infinity then $\sigma_{ess}(L) = \sigma_{ess}(L+C)$.\\

\noindent \textit{Proof.} For some (and hence every) $z$ in the resolvent of $L, R = 
(L - zI)^{-1}$ is a bounded operator from $L_{\beta}^{2}$ to $\mathcal{H}_{B}$ and 
hence, for some constant $k$,
\begin{center}
 $||Rf||_{\mathcal{H}_{B}} \leq k||f||_{L_{\beta}^{2}}$.
\end{center}
Choose an exhaustion $\lbrace M_{n} \rbrace$ of $M$ and cutoff functions $u_{n}$ with 
supp $u_{n} \subset M_{n}$, $u_{n} \equiv 1$ on $M_{n-1}$ and $|\nabla u_{n} | 
\rightarrow 0$ as $n \rightarrow \infty$.  By Rellich's lemma, $D_{n} = u_{n}CR$
is compact on $L_{\beta}^{2}(M)$.\\

\noindent \textit{Claim.} $C$ is relatively compact with respect to $L$; i.e., 
$D = CR$ is compact on $L_{\beta}^{2}(M)$.  This follows from the inequality:
\begin{align*}
 &||(D-D_{n})f||^{2}_{L_{\beta}^{2}(M)} = ||(1-u_{n})CRf||^{2}_{L_{\beta}^{2}(M)}&\\
 &\hspace*{1.12 in} \leq \int\limits_{M \setminus M_{n-1}} |CRf|^{2}\cosh^{2}\beta r \; dV&\\
 &\hspace*{1.12 in} \leq k'\tau(r_{n})||Rf||_{L_{\beta}^{1,2}(M)}^{2}&\\
 &\hspace*{1.12 in} \leq k''\tau(r_{n})||f||_{L_{\beta}^{2}(M)}^{2}&\\
\end{align*}\vspace*{-0.5 in}\\
\noindent which shows that $D_{n}$ converges to $D$ in norm.  The result now follows from\\

\newcounter{foo}
\setcounter{foo}{1}

\noindent \textbf{Theorem. } (Weyl, cf. [RS] Corollary 2, IV p. 113) If $L$ is
as in Proposition 5.2 and $\exists z \in \mathbb{C}$ such that $C(L - zI)^{-1}$ is
compact, then $\sigma_{ess}(L) = \sigma_{ess}(L+C)$.\\

The prototype of ordinary differential operators which arise in the Donnelly 
decomposition is 
\begin{equation}\tag{5.3}
 \mathcal{D}f = -\dfrac{d^{2}f}{dx^{2}} - 2\gamma\dfrac{df}{dx} + c(x)f\\
\end{equation}
\noindent where $c(x)$ is rapidly decaying, $\gamma > 0$ is constant, and 
$f \in L^{2}(R^{+},e^{2\gamma x}dx)$.\\ 

\noindent \textbf{Lemma 5.4 } Up to compact perturbation, $\mathcal{D}$
as defined in (5.3) is unitarily equivalent to the multiplication operator, 
$(\mathcal{M}f)(x) = (x^{2} + \gamma^{2})f(x)$ acting on $L^{2}(R^{+},dx)$.  
Therefore, $\sigma_{ess}(\mathcal{D}) = \sigma(\mathcal{D}) = [\gamma^{2},\infty)$.\\

To prove this, recall that the change of dependent variable, $f = e^{-\gamma x}k = Uk$,
defined a unitary transformation $U$ from $L^{2}(R^{+},dx)$ to 
$L^{2}(R^{+},e^{2\gamma x}dx)$ under which
\begin{center}
 $\mathcal{D}_{1}k := (U^{-1}\mathcal{D}U)k = -\dfrac{d^{2}k}{dx^{2}} 
+ (\gamma^{2} + c)k,$\\
\end{center}

\noindent so that $\mathcal{D}$ is unitarily equivalent to $\mathcal{D}_{1}$ acting 
on ordinary $L^{2}(R^{+},dx)$.  From proposition 5.2, $\mathcal{D}_{1}$ has the 
same essential spectrum as
\begin{center}
$\mathcal{D}_{2}k = -\dfrac{d^{2}k}{dx^{2}} + \gamma^{2}k$
\end{center}
and, by Fourier transformation, $\mathcal{D}_{2}$ is unitarily equivalent to the 
multiplication operator $\mathcal{M}$ in the lemma.  The conclusion about the 
spectrum is immediate.\\

We are now ready to prove the main result of this section, Proposition 5.1, 
concerning the spectrum of the Laplacian (on 0-forms and 1-forms) and of the
Floer operator (at the zero connection).  Recall that we have already shown, in
Section 4, the absence of eigenvalues, so that it suffices to obtain the results
for the essential spectrum.\\

We use $d_{s}$ and $d_{s}^{*}$ to denote, respectively, exterior differentiation 
and its $L^{2}$ adjoint on $S^{2}$.  Let $\Delta_{s}$ denote the Laplacian on $S^{2}$.
 For notational simplicity, we write $g = \sinh r$ and $w = \cosh^{2}\beta r$.
 Sometimes, for clarity, we use the subscriptp 0 or 1 on the Laplacian to distinguish
the domain as functions or 1-forms.\\

On $\mathbb{H}^{3}$, the formulas for the weighted Laplacian are: for 
$\varphi \in \wedge^{0}$,
\begin{center}
$\Delta_{0}\varphi = g^{-2}\Delta_{s}\varphi - 
g^{-2}w^{-1}\dfrac{\partial}{\partial r}(g^{2}w\dfrac{\partial \varphi}{\partial r})$
\end{center}
\noindent  and for $a = a_{1} + a_{2}dr \in \wedge^{1}$,
\begin{center}
 $\Delta_{1}a = g^{-2}\Delta_{s}a - w^{-1}\dfrac{\partial}{\partial r}
$\begin{LARGE}(\end{LARGE} $w\dfrac{\partial a_{1}}{\partial r}$ \begin{LARGE}$)$\end{LARGE}$- \dfrac{\partial}{\partial r}$
\begin{LARGE}(\end{LARGE}$g^{-2}w^{-1}
\dfrac{\partial}{\partial r}(g^{2}wa_{2})$\begin{LARGE}$)$\end{LARGE}$\wedge \; dr$\\
\hspace*{-1.5 in} $-2g^{-1}\dfrac{\partial g}{\partial r}(d_{s}a_{2}+g^{-2}d_{s}^{*}
a_{1} \wedge dr)$
\end{center}

To prove the proposition, we separate variables and expand any $p$-form in 
eigenfunctions on $S^{2}$.  Then, every $\varphi \in \wedge^{0}$ is a sum
of terms of the form $h_{0} \tau_{0}$ (with $\tau_{0}$ an eigenfunction on 
$S^{2}$).  Using a Hodge decomposition, one sees that every 1-form on 
$\mathbb{H}^{3}$ is a sum of three terms (corresponding to the eigenvalue 
$\lambda$ of $\Delta_{s}$) of the form:
\begin{equation}\tag{5.5}
 h_{1}(r)\tau_{1} + h_{2}(r)\tau_{2}dr + (h_{3}(r)d_{s}\tau_{3} + h_{4}(r)\tau_{3}dr)
\end{equation}
where $\tau_{1}$ is a co-closed eigen 1-form, and $\tau_{2}$ and $\tau_{3}$
are eigenfunctions on $S^{2}$ (with $\tau_{2} =$ constant corresponding to 
$\lambda = 0$ and $\tau_{3}$ occuring only if $\lambda \neq 0$).  This 
decomposition into three types is orthogonal and is preserved both by the 
Laplacian and the Floer operator.\\

Note that the forms under consideration will be in $L_{\beta}^{2}(\mathbb{H}^{3})$
if and only if 
\begin{equation}\tag{5.6}
||h_{i}||_{2,\gamma_{i}}^{2} = \int\limits_{0}^{\infty}h_{i}^{2}(r)\gamma_{i}(r) \; dr 
< \infty \hspace*{0.5 in} i = 0, 1, 2, 3, 4
\end{equation}

\noindent where $\gamma_{0} = \gamma_{2} = \gamma_{4} = g^{2}w$ and 
$\gamma_{1} = \gamma_{3} = w$.\\

In this context, the Laplacian defines ordinary differential operators
on the spaces in (5.6) as follows:
\begin{align*}
&(\text{i})& 
 &(\mathcal{D}_{0}h_{0})\tau_{0} = \Delta_{0}(h_{0}\tau_{0})
 = (-g^{-2}w^{-1}\dfrac{d}{dr}(g^{2}w\frac{dh_{0}}{dr})+ \lambda g^{-2} h_{0})\tau_{0}&
\\
&(\text{ii})&
&
 (\mathcal{D}_{1}h_{1})\tau_{1} = \Delta_{1}(h_{1}\tau_{1}) 
 = (-w^{-1}\dfrac{d}{dr}(w\frac{dh_{1}}{dr}) + \lambda g^{-2}h_{1})\tau_{1}
&
\\
&(\text{iii})&
&
 (\mathcal{D}_{2}h_{2})\tau_{2} \; dr = \Delta_{1}(h_{2}\tau_{2} \; dr)
 = (-\dfrac{d}{dr}(g^{-2}w^{-1}\frac{d}{dr}(g^{2}wh_{2}))+ \lambda g^{-2}h_{2})\tau_{2} \; dr
&
\\
&(\text{iv})&
&
 \mathcal{D}_{3}(h_{3},h_{4}) = (\mathcal{D}_{1}h_{3} + 2g^{-1}\dfrac{dg}{dr}h_{4},
\mathcal{D}_{2}h_{4} + 2g^{-3}\dfrac{dg}{dr}h_{3}).
&
\end{align*}

\indent The explanation for (iv) is that
\begin{center}
$\Delta_{1}(h_{3}d_{s}\tau_{3}+h_{4}\tau_{3} \; dr) = 
(\mathcal{D}_{1}h_{3} + 2g^{-1}\dfrac{dg}{dr}h_{4})d_{s}\tau_{3}
 + (\mathcal{D}_{2}h_{4} + 2g^{-3}\dfrac{dg}{dr}h_{3})\tau_{3} \; dr$
\end{center}

\noindent Note that $\mathcal{D}_{0}$ and $\mathcal{D}_{2}$ are operators
on $L^{2}(R^{+}, g^{2}wdr), \mathcal{D}_{1}$ is an operator on $L^{2}(R^{+},wdr)$
and $\mathcal{D}_{3}$ acts on $L^{2}(R^{+},wdr) \times L^{2}(R^{+},g^{2}wdr)$. \\

It follows immediately that $\mathcal{D}_{1}$ is of the form (5.3) with 
$\gamma = \beta$.  Hence, $\sigma(\mathcal{D}_{1}) = [\beta^{2}, \infty)$.\\

Using Proposition 5.2, we may replace $g^{2}$ by $e^{2r}$ and $w$ by $e^{2\beta r}$
 in $\mathcal{D}_{0}$ and $\mathcal{D}_{2}$ without changing the spectra.  This
 gives us an operator $\mathcal{\tilde{D}}$ acting on 
$h \in L^{2}(R^{+},e^{2(1+\beta)r}dr)$ given by:

\begin{center}
 $\mathcal{\tilde{D}}h = -\dfrac{d}{dr}(e^{-2(1+\beta)r}\dfrac{d}{dr}(e^{2(1+\beta)r}h)) 
+ \lambda e^{-2r}h$,
\end{center}

\noindent which is of the form (5.3) with $\gamma = 1 + \beta$.  Hence, 
$\sigma(\mathcal{D}_{0}) = \sigma(\mathcal{D}_{2}) = [(1+\beta)^{2},\infty).$\\

Finally, in the third case, again using Proposition 3.14, it suffices to consider
the spectrum of
\begin{center}
 $\mathcal{D}_{3}'(h_{3},h_{4}) = (\mathcal{D}_{1}h_{3} + 2h_{4}, \mathcal{D}_{2}h_{4} + e^{-2r}h_{3})$
\end{center}
acting on pairs $(h_{3},h_{4}) \in L^{2}(R^{+},e^{2\beta r}dr) \times L^{2}(R^{+},e^{2(1+\beta)r}dr)$.
Making the unitary change of variable $(h_{3},h_{4}) = (e^{-\beta r}k_{3}, e^{-(1+\beta)r}k_{4})$
gives the operator $\mathcal{D}_{3}'' = U^{-1}\mathcal{D}_{3}'U$ where
\begin{center}
 $\mathcal{D}_{3}''(k_{3},k_{4}) = (\dfrac{-d^{2}k_{3}}{dr^{2}}+\beta^{2}k_{3} 
+ 2e^{-r}k_{4}, \dfrac{-d^{2}k_{4}}{dr^{2}}+(1+\beta)^{2}k_{4} + 2e^{-r}k_{3})$
\end{center}

\noindent on pairs $(k_{3},k_{4}) \in L^{2}(R^{+},dr) \times L^{2}(R^{+},dr)$.
Another application of Proposition 5.2 shows that the spectrum of $\mathcal{D}_{3}$
is 
\begin{center}
 $\sigma(\mathcal{D}_{3}) = [\beta^{2},\infty) \cup [(1+\beta^{2}),\infty)
 = [\beta^{2},\infty)$.
\end{center}

\noindent Since up to compact perturbation, $\Delta_{p}$ is unitarily equivalent
to sums of the above operators, we have demonstrated (a) and (b) of proposition
(5.1).\\

To prove (c), we show that $\delta_{0}\delta_{0}^{\dagger}$ differs from the 
Laplacian on pairs by a compact perturbation.  Recall (3.16) that 
$\delta_{0}\delta_{0}^{\dagger}(\psi,b)=(\Delta_{0}\psi,\Delta_{1}b) + (0,T_{1}b) $\\
\noindent where
\begin{center}
 $T_{1}b = -2\beta\tanh \beta r\lbrace*(dr \wedge *db) + d^{*}(b \wedge dr) \rbrace$
\end{center}

Evaluating $T_{1}$ on each of the three types of one forms occurring in (5.5), 
we find that $T_{1}b = 0$ on the subspaces spanned by $\tau_{1}$ and $\tau_{2}dr$.
For $b$ of the third kind, $T_{1}$ gives rise to the operator
\begin{center}
 $\mathcal{T}(h_{3},h_{4}) = -2\beta\tanh \beta r(h_{4}, \lambda g^{-2} h_{3})$
\end{center}
since $T_{1}(h_{3}d_{s}\tau_{3} + h_{4}\tau_{3}\;dr) = -2\beta\tanh \beta r(h_{4}\;d_{s}\tau_{3} + 
\lambda g^{-2} h_{3}\tau_{3}\;dr)$.  As before, $(h_{3},h_{4}) \in L^{2}(R^{+}, e^{2\beta r}dr)
\times L^{2}(R^{+},e^{2(1+\beta)r}dr)$.  The unitary change of dependent variable 
$(h_{3},h_{4}) = (e^{-\beta r}k_{3}, e^{-(1+\beta)r}k_{4})$ gives a unitary equivalence
 of $\mathcal{T}$ with
\begin{center}
 $\mathcal{\hat{T}}(k_{3},k_{4}) = (U^{-1}\mathcal{T}U)(k_{3},k_{4}) = -2\beta \tanh \beta r
(e^{-r}k_{4}, \lambda e^{r}g^{-2}k_{3})$
\end{center}
acting on $L^{2}(R^{+},dr) \times L^{2}(R^{+},dr)$.\\

\noindent It follows that $T_{1}$ contributes a compact perturbation to $\Delta_{1}$,
and hence, by Prop 5.2 does not change the spectrum from which (c) of proposition
5.1 follows.

\section{The Existence Theorem}
\indent We recall the \\

\noindent \textbf{Raleigh Quotient Theorem.} \hspace*{-0.1 in} (cf [Da] Theorem 4.3.1, p. 78) If $L$ is a self-adjoint
operator on a Hilbert space $\mathcal{H}$, then $(Lf,f)_{\mathcal{H}} \geq c ||f||_{\mathcal{H}}^{2}$
for all $f$ in the domain of $L$, if and only if the spectrum $\sigma(L) \subset [c, \infty)$.\\

\indent Then, letting $\Delta = (\Delta_{0}, \Delta_{1})$ denote the Laplacian on 
configuration pairs, the information in Proposition 5.1 may be translated 
(using the Raleigh-Quotient Theorem) into the inequalities:\vspace*{0.20 in}\\
(6.1)\vspace*{-0.45 in}
\begin{align*}
&(a',b')& &\beta^{2}||\eta||^{2}_{2,\beta} \leq (\Delta\eta,\eta)_{\beta}& \\
&(c')& &\beta^{2}||\eta||_{2}^{2,\beta} \leq 
(\delta_{0}\delta_{0}^{\dagger}\eta,\eta)_{\beta} = (\delta_{0}^{\dagger}\eta,\delta_{0}^{\dagger}\eta)_{\beta}&
\end{align*}

Next, we show that the scalar operators are invertible on $\mathcal{H}_{0}$.  (Recall
that $\mathcal{H}_{0} = L^{2,2}_{\beta}$.)\\

\noindent \textbf{Proposition 6.2.}  There is a constant $\kappa > 0$ such that
\begin{center}
 $\kappa||\eta||_{\mathcal{H}_{0}} \leq ||\delta_{0} \delta_{0}^{\dagger}\eta||_{2,\beta}$
\end{center}
and $\delta_{0}\delta_{0}^{\dagger}$ is invertible.\\

\textit{Proof.}  First, we note that since $\delta_{0}\delta_{0}^{\dagger}$ is a self-adjoint
operator defined on $\mathcal{H}_{0}$, the inequality implies that the co-kernel is zero
and hence, $\delta_{0}\delta_{0}^{\dagger}$ is not only injective but surjective with
closed range, and hence invertible.\\

To prove the inequality, we recall the Bochner-Lichnerowicz-Weitzenbock formula 
at a point, for the standard Laplacian $\Delta_{p}^{0} = dd^{*} + d^{*}d$ ([FU]):
\begin{equation}\tag{6.3}
 -\nabla^{2}\omega = \Delta_{p}^{0}\omega - \text{Ric}_{p}(\omega,\cdot), \hspace*{0.25 in} 
\omega \in \wedge^{p}, p = 0, 1
\end{equation}
and Ric$_{0} = 0$.\\

Letting $\nabla^{\dagger}\omega = \cosh^{-2}$\hspace*{-0.1pt}$\beta r \nabla ((\cosh^{2}\beta r) \omega)$, 
and, noting that\\
\begin{equation}\tag{6.4}\Delta_{p}\omega = (dd^{\dagger} + d^{\dagger}d)\omega = 
\Delta_{p}^{0}\omega + 2\beta\tanh \beta r(\ast(\;dr \wedge \ast \; d\omega)) + 2\beta d(\ast (\tanh \beta r \; dr \wedge \ast \omega))
\end{equation}
we obtain\\

$
 -\nabla^{\dagger}\nabla\omega = \Delta_{p}\omega - \text{Ric}_{p}(\omega, \cdot) 
 - 2 \beta\tanh \beta r(\ast(\;dr \wedge \ast \; d\omega))$\\
\hspace*{1.05 in}$- 2\beta d(\ast (\tanh \beta r \; dr \wedge \ast \omega))
$

An integration by parts and use of (6.1) gives
\begin{center}
 $|| \nabla \omega||_{2,\beta}^{2} \leq (\Delta_{p}\omega, \omega) + c|| \nabla \omega ||_{2,\beta}||\omega||_{3,\beta} + c'||\omega||^{2}_{2,\beta}$\\
 \hspace*{0.06 in} $\leq (1+C(\epsilon))(\Delta_{p}\omega,\omega)+ \epsilon||\nabla \omega||_{2,\beta}^{2}$
\end{center}

\noindent from which it follows that
\begin{equation}\tag{6.5}
 ||\omega||_{L_{\beta}^{1,2}} \leq C||\Delta_{p} \omega ||_{2,\beta}
\end{equation}

\indent Using the Weitzenbock formula once again,
\begin{center}
 $|| \nabla^{2} \omega ||_{2,\beta}^{2} \leq || \Delta_{p}\omega ||_{2,\beta}^{2} +
 C'|| \omega ||_{L_{\beta}^{1,2}}^{2} \leq C''|| \Delta_{p} \omega ||_{2,\beta}^{2}$
\end{center}
\noindent which gives
\begin{equation}\tag{6.6}
 || \omega ||_{L_{\beta}^{2,2}} \leq C|| \Delta_{p} \omega ||_{2,\beta}, \; \; \; \omega \in \wedge^{p} \; \; p = 0, 1.
\end{equation}

\noindent Next, recall again (3.16) that on pairs,
\begin{center}
 $\delta_{0}\delta_{0}^{\dagger}\eta = \Delta \eta + T \eta $ where $T = (0,T_{1})$
\end{center}
\noindent and $T_{1}$ is defined by (3.12).  Using (6.6), (6.1c'), Proposition 3.18 and Sobolev's inequality
\begin{align*}
 &||\eta||_{\mathcal{H}_{0}}\hspace*{-0.45 in}& &\leq C|| \Delta \eta ||_{2,\beta} \leq C (||\delta_{0}\delta_{0}^{\dagger}\eta||_{2,\beta} + 
||T_{1}b||_{2,\beta})&\\
&& &\leq C||\delta_{0}\delta_{0}^{\dagger}\eta||_{2,\beta} 
+ C'||b||_{L_{\beta}^{1,2}}&\\
&&&\leq C||\delta_{0}\delta_{0}^{\dagger}\eta||_{2,\beta} + C(\epsilon)||\eta||_{2,\beta}
 + \epsilon||\eta||_{\mathcal{H}_{0}}&\\
&&&\leq C'||\delta_{0}\delta_{0}^{\dagger}\eta||_{2,\beta} + \epsilon ||\eta||_{\mathcal{H}_{0}}&
\end{align*}

\noindent Absorbing the term $\epsilon||\eta||_{\mathcal{H}_{0}}$, proves the
inequality of Proposition 6.2.\\

Finally, letting $\mathcal{L} = \delta\delta^{\dagger}$, (recall the definitions
of $\delta$ and $\delta^{\dagger}$ from (3.4) and (3.7)) we are now ready to 
prove the main estimate of this paper.\\

\noindent \textbf{Theorem 6.7.} There is a constant $\alpha_{1} > 0$ such that
\begin{center}
 $\alpha_{1} || \eta ||_{\mathcal{H}_{A_{0}}} \leq || \mathcal{L} \eta ||_{2,\beta}$
and $\mathcal{L}$ is invertible.
\end{center}

\noindent We notice that, as before, $\mathcal{L}$ is self-adjoint on its domain
$\mathcal{H}_{A_{0}}$ and hence, the inequality shows invertibility.\\

To prove the theorem, we compare the various operators evaluated at $A_{0}$ with
their scalar analogues to show that the lower bounds on spectra do not decrease.\\

We note first that from (3.13) we have that $\Delta_{A_{0}} = \Delta + S$ with $S$ rapidly decaying at infinity.  Propositions (3.18) and (5.2) then imply
\begin{equation}\tag{6.8}
 \sigma (\Delta_{A_{0}}) = \sigma_{ess}(\Delta_{A_{0}}) \subseteq [\beta^{2},\infty)
\end{equation}
Similarly, we recall (3.17) $\delta_{A_{0}}\delta_{A_{0}}^{\dagger} = \delta_{0}\delta_{0}^{\dagger} + R + S$.\\

\noindent Again Propositions (3.18) and (5.2) imply 
\begin{equation}\tag{6.9}
 \sigma(\delta_{A_{0}}\delta_{A_{0}}^{\dagger}) = \sigma_{ess}(\delta_{A_{0}}\delta_{A_{0}}^{\dagger}) \subseteq [\beta^{2},\infty).
\end{equation}

\noindent Finally, recalling (3.8), $\delta\delta^{\dagger} = \delta_{A_{0}}\delta_{A_{0}} - (ad\Phi)^{2} + E$, Propositions (3.18) and (5.2), together with the observation that $-(ad\Phi_{0})^{2}$ is a non-negative operator, result in
\begin{equation}\tag{6.10}
 \sigma(\delta\delta^{\dagger}) = \sigma_{ess}(\delta\delta^{\dagger}) \subseteq [\beta^{2},\infty)
\end{equation}

Using the Raleigh-Quotient Theorem,
\begin{equation}\tag{6.11}
 \begin{cases}
  \beta^{2} || \eta ||_{2,\beta}^{2} \leq (\Delta_{A_{0}}\eta, \eta)_{\beta}\\
  \kappa||\eta||_{2,\beta}^{2} \leq (\delta \delta^{\dagger}\eta,\eta)_{\beta} \text{ for some }\kappa \geq \beta^{2}.
 \end{cases}
\end{equation}

To prove the inequality in the theorem, we use the Weitzenbock formula at $A_{0}$, 
with $\beta = 0$ to obtain 
\begin{equation}\tag{6.12}
 d_{A_{0}}^{*}d_{A_{0}} + d_{A_{0}}d_{A_{0}}^{*} = - \nabla_{A_{0}}^{2}+\lbrace Ric, \cdot\rbrace
+ \lbrace F_{A_{0}}, \cdot \rbrace.
\end{equation}

For the weighted operators, with $\beta > 0$, we find 
\begin{equation*}
 \Delta_{A_{0}} = d_{A_{0}}^{\dagger}d_{A_{0}} + d_{A_{0}}d_{A_{0}}^{\dagger} 
= - \nabla_{A_{0}}^{\dagger}\nabla_{A_{0}} + Q_{1}\vspace*{-0.1in}
\end{equation*}\vspace*{-0.35 in}
\begin{flushleft}
(6.13) 
\end{flushleft}
\vspace*{-0.222 in}
\hspace*{1.125 in} $\delta\delta^{\dagger}= - \nabla_{A_{0}}^{\dagger}\nabla_{A_{0}} + Q_{2}$\\

\vspace*{-0.1 in}\noindent where $|Q_{i} \eta| \leq c(|\eta|+|\nabla\eta|)$, $i = 1, 2$\\

As before,
\begin{flushleft}
(6.14) 
\end{flushleft}
\vspace*{-0.35 in}
\hspace*{0.75 in}$|| \nabla_{A_{0}} \eta||_{2,\beta}^{2} \leq (\delta\delta^{\dagger}\eta,\eta)_{\beta} 
+ (Q_{2}\eta,\eta)_{\beta}$ \\
\hspace*{1.47 in} $\leq (\delta\delta^{\dagger}\eta,\eta)_{\beta} + 
\epsilon|| \nabla_{A_{0}} \eta ||_{2,\beta}^{2} + C(\epsilon)|| \eta ||_{2,\beta}^{2}$\\

and therefore, from (6.11),
\begin{equation}\tag{6.15}
 ||\eta||_{2,\beta} + || \nabla_{A_{0}} \eta||_{2,\beta} \leq C'||\delta \delta^{\dagger}\eta||_{2,\beta}
\end{equation}

\indent Using (6.6), (6.1c), Proposition 3.18 and Sobolev's inequality prove the theorem.\\

Note that because of the explicit knowledge about the Chakrabarti monopole,
the decay of the approximate monopole is known for $\mathbb{H}^{3}$ and also 
that the basic estimate, Theorem (6.7) holds without any assumption that
$Lc_{0} = d_{A_{0}}\Phi_{0} - *F_{A_{0}}$ is small.  However, this condition
will be needed later in applying the Implicit Function Theorem.  The restriction
$\beta < m$, required to make $Lc_{0}$ small, will also be used later.\\

\noindent \textbf{Corollary 6.16. } If $||\nu||_{6,\beta}$ is sufficiently small, then
\begin{center}
 $\mathcal{L}_{\nu} = \mathcal{L} + \nu\#\delta^{\dagger}$
\end{center}
is invertible.\vspace*{0.25 in} \\ 
\noindent \textit{Proof. } We use a weighted version of Sobolev's inequality which
says that, for $2 < p \leq 6$,
\begin{equation}\tag{6.17}
 ||\eta||_{p,\beta} \leq C(|| \nabla_{A_{0}} \eta||_{2,\beta} + ||\eta||_{2,\beta})
\end{equation}
This inequality implies that 
\begin{center}
 $||\eta||_{p,\beta} \leq C||\eta||_{\mathcal{H}_{A_{0}}}$ and 
 $|| \zeta^{\dagger} \eta||_{p,\beta} \leq C'||\eta||_{\mathcal{H}_{A_{0}}}$.
\end{center}

\noindent Using the Holder's inequality, with $\alpha_{2} = \text{max}(C,C')$,\vspace*{0.25 in}\\
(6.18)\vspace*{-0.5 in}
\begin{center}
 \hspace*{0.45 in}$||v\#\delta^{\dagger}\eta||_{2,\beta} \leq ||v||_{6,\beta}||\delta^{\dagger}\eta||_{3,\beta} 
\leq \alpha_{2}||v||_{6,\beta}||\eta||_{\mathcal{H}_{A_{0}}}
\leq \alpha_{2}^{2}||v||_{\mathcal{H}_{A_{0}}}||\eta||_{\mathcal{H}_{A_{0}}}$\vspace*{0.10 in}\\
 \hspace*{-2.00 in}$||\delta^{\dagger}\tau \# \delta^{\dagger} \eta||_{2,\beta} \leq \alpha_{2}^{2}||\tau||_{\mathcal{H}_{A_{0}}}||\eta||_{\mathcal{H}_{A_{0}}}$
\end{center}

Using Theorem 6.7,
\begin{equation}\tag{6.19}
 \alpha_{1} || \eta ||_{\mathcal{H}_{A_{0}}} \leq || \mathcal{L}_{\nu}\eta ||_{2,\beta} + ||\nu \# \delta^{\dagger}\eta||_{2,\beta},
\end{equation}

\noindent which combined with (6.18) gives, for $||\nu||_{6, \beta}$ sufficiently small,
\begin{equation}\tag{6.20}
 \alpha'||\eta||_{\mathcal{H}_{A_{0}}} \leq || \mathcal{L}_{\nu}\eta ||_{2,\beta}
\end{equation}

\noindent Invertibility of $\mathcal{L}_{\nu}$, for $||\nu||_{6,\beta}$ small, follows
from the fact that $\mathcal{L}_{0}$ is invertible and 
$||\mathcal{L}_{0}-\mathcal{L}_{v}||_{2,\beta} \leq C||\nu||_{6,\beta}$.\\

 Recall that we have seen in section 3, that to solve the Bogomolny monopole equation
\begin{equation}\tag{1.1}
 Lc = L(\Phi, A) = d_{A}\Phi - *F_{A} =  0
\end{equation}
it suffices to find a solution $\eta = (\psi,b)$ of 
\begin{equation}\tag{3.6}
 \mathscr{L} \eta = \delta \delta^{\dagger} \eta + \delta^{\dagger}\eta \# \delta^{\dagger}\eta = G_{0}
\end{equation}
for $G_{0}$ and $\delta^{\dagger} \eta \# \delta^{\dagger} \eta$ sufficiently small.\\

\noindent \textbf{\underline{Theorem.} } The Bogomolny monopole equation (1.1) has a solution \\$c = (\Phi,A)$ with given magnetic charge $k \in Z$ and (arbitrary) mass $m \in \mathbb{R}^{+}$.\vspace*{0.25 in}\\
\textit{\underline{Proof.} } Following [FU], we apply the continuity method to the equation 
\begin{equation}\tag{\textasteriskcentered}
 \mathscr{L}\eta_{t} = \delta\delta^{\dagger}\eta_{t}+\delta^{\dagger}\eta_{t} \# \delta^{\dagger}\eta_{t} = tG_{0}, \; \;  0 \leq t \leq 1.
\end{equation}

To that end, let $\lambda < \alpha_{1} / 4\alpha_{2}^{2}$ where
$\alpha_{1}$ and $\alpha_{2}$ are the constants occurring in the inequalities 
(6.7) and (6.18).  Also, assume that $||G_{0}||_{2,\beta} \leq \alpha_{1}\lambda/4$.

Let 
\begin{center}
 $\Omega = \lbrace \eta \in \mathcal{H}_{A_{0}}: ||\eta||_{\mathcal{H}_{A_{0}}} \leq \lambda \rbrace$
\end{center}

\noindent and 
\begin{center}
 $J = \lbrace t \in [0,1]$ equation (\textasteriskcentered) has a solution in $\Omega\rbrace$
\end{center}

We show that $J$ is non-empty, open and closed.  Clearly, $t = 0$ belongs to 
$J$ since $\eta \equiv 0$ is the unique solution.\\
\indent To show that $J$ is open, let $t_{0} \in J$, with $\eta_{0}$ the corresponding solution of (\textasteriskcentered)
belonging to $\Omega$.  The linearized operator at $\eta_{0}$ is $\mathcal{L}_{\nu}$
with $\nu = 2\delta^{\dagger}\eta_{0}$.  From (6.16) and (6.17) at 
$2\delta^{\dagger}\eta_{0}$ and the choice of $\lambda$ above, one has $||\nu\#\delta^{\dagger}\eta||_{2,\beta}
\leq \frac{\alpha_{1}}{2} ||\eta ||_{\mathcal{H}_{A_{0}}}$.  It follows from
(6.18) that (6.20) holds with $\alpha' = \frac{\alpha_{1}}{2}$ and 
$\mathcal{L}_{\nu}$ is invertible.

From the Implicit Function Theorem, we conclude that (\textasteriskcentered) has a solution $\eta_{t}$
for $t$ sufficiently close to $t_{0}$, and $||\eta_{t} -\eta_{0}||_{\mathcal{H}_{A_{0}}} 
< \epsilon$, for $\epsilon$ sufficiently small.  Estimating again (as in 6.18),
and using the fact that $\eta_{0} \in \Omega$, we find, from (6.7)\\

 \hspace*{0.65 in} $\alpha_{1}|| \eta_{0}||_{\mathcal{H}_{A_{0}}} \leq 
||t_{0}G_{0}||_{2,\beta} + ||\delta^{\dagger}\eta_{0}\#\delta^{\dagger}\eta_{0}
||_{2,\beta}$\vspace*{-0.25 in}

\begin{equation}\tag{6.21}
\leq ||t_{0}G_{0}||_{2,\beta} + \dfrac{\alpha_{1}}{4}||\eta_{0}||_{\mathcal{H}_{A_{0}}}
\end{equation}

\noindent Using the bound on $G_{0}$ from Proposition 2.7, we find
\begin{equation}\tag{6.22}
 ||\eta_{0}||_{\mathcal{H}_{A_{0}}} \leq \dfrac{1}{3}\lambda
\end{equation}

\noindent so that for $\eta_{t}$ sufficiently close to $\eta_{0}$, which will be the
case if $t$ is close to $t_{0}$, we see that
\begin{equation}\tag{6.23}
 ||\eta_{t}||_{\mathcal{H}_{A_{0}}} \leq \lambda.
\end{equation}

\noindent Therefore, $J$ is open.

To prove that $J$ is closed, let $t_{n} \in J$ converge to $t_{0}$.  Then, for each
$n$, let $\eta_{n}$ be the solution of (\textasteriskcentered) corresponding to $t_{n}$.  Since
$||\eta_{n}||_{\mathcal{H_{A_{0}}}} \leq \lambda$, a subsequence converges weakly
in $\mathcal{H}_{\mathcal{A}_{0}}$ to $\eta_{0}$ and by lower semi-continuity with respect 
to weak convergence, $||\eta_{0}||_{\mathcal{H}_{\mathcal{A}_{0}}} \leq \lambda$.  We claim
that $ \mathscr{L}\eta_{0} = t_{0}G_{0}$.  It suffices to show this on any compact
subdomain.  The linear term $\delta\delta^{\dagger}\eta_{n}$ converges weakly
to $\delta\delta^{\dagger}\eta_{0}$.  By Sobolev embedding $\delta^{\dagger}\eta_{n}$
converges strongly to $\delta^{\dagger}\eta_{0}$ in $L_{\beta}^{p}$ for $p < 6$ and 
therefore, $\delta^{\dagger}\eta_{n}\#\delta^{\dagger}\eta_{n}$ converges strongly
to $\delta^{\dagger}\eta_{0}\#\delta^{\dagger}\eta_{0}$ in $L_{\beta}^{2}$, since
$||\delta^{\dagger}\eta_{0}\#\delta^{\dagger}\eta_{0}||_{2,\beta} \leq ||\delta^{\dagger}\eta_{0}||_{4,\beta}^{2}$
and Lebesgue dominated convergence is applicable.  It follows that $ \mathscr{L}\eta_{n}$
converges weakly to $\mathscr{L}\eta_{0}$ which is a solution of the equation, as 
desired.  This shows that J is closed, and completes the proof of our main theorem.

\newpage
\thispagestyle{plain}
\begin{center}
R\sc{eferences}
\end{center}

\noindent \hspace*{-0.70 in}\text{[A]  } \hspace*{0.38 in}\text{Atiyah, M.F.,} \textit{Magnetic monopoles in hyperbolic spaces: Vector bundles 
on algebraic varieties}, Tata Institute of Fundamental Research, Bombay (1984),
1-33.\\
 \hspace*{-0.75 in}\text{[Br] } \hspace*{0.39 in}\text{Braam, P.J.,} \textit{Magnetic monopoles on three-manifolds},
\text{J. Diff. Geom. } 30 (1989), 425-464.\\ 
 \hspace*{-0.75 in}\text{[Bo] } \hspace*{0.35 in}\text{Bogomolny, E.B.,} \textit{The stability of classical solutions}, Sov. J. Nucl. Phys. 24 (1976), 449.\\  
\hspace*{-0.75 in}\text{[C]  } \hspace*{0.38 in} \text{Chakrabarti, A.,} \textit{Spherically and axially symmetric
SU(n) instanton chains with monopole limits}, Nucl. Phys. B 248 (1984), 209-252.\\
 \hspace*{-0.75 in}\text{[D]  } \hspace*{.42 in}\text{Donaldson, S.K.,} \textit{Anti-self-dual Yang Mills connections over complex algebraic surfaces and stable vector bundles},
Proc. Lond. Math. Soc. 30 (1985), 1-26.  \\
 \hspace*{-0.75 in}\text{[Da] } \hspace*{.32 in}\text{Davies, E.B.,} \textit{Spectral Theory and Differential Operators},
Cambridge Univ. Press (1995).  \\
 \hspace*{-0.75 in}\text{[Di] }\hspace*{.38 in}\text{ Dirac, P.A.M.,} \textit{Proc Roy Soc}, A 133, 60 (1931) \\
  \hspace*{-0.75 in}\text{[Do] } \hspace*{.34 in}\text{Donnelly, H.,} \textit{Differential Form Spectrum of
Hyperbolic space}, Manuscripta Math. 33 (1981), 365-385. \\
 \hspace*{-0.75 in}\text{[Du] }\hspace*{.4 in}\text{Durenard, E.,} \textit{Mayer Vietoris Result for 
Monopoles}, Harvard Ph. D. thesis (1995).\\
  \hspace*{-0.75 in}\text{[E]  } \hspace*{.44 in}\text{Ernst, K.D.,} \textit{The ends of the Monopole Moduli space over }$\mathbb{R}^{3}\#$\textit{(Homology sphere): I and II}, The Floer Memorial Volume, Birkhauser (1995), 355-434.  \\
 \hspace*{-0.75 in}\text{[F1] } \hspace*{.38 in}\text{Floer, A.,} \textit{Monopoles on asymptotically 
Euclidean} 3\textit{-manifolds.}, Bull. Amer. Math. Soc. 16 (1987), 125-127.\\
 \hspace*{-0.75 in}\text{[F2] } \hspace*{.36 in}\text{Floer, A.,} \textit{Monopoles on asymptotically flat manifolds}, preprint (1987), The Floer Memorial Volume, Birkhauser (1995), 3-42.\\
 \hspace*{-0.75 in}\text{[FHP1]} \hspace*{.2 in}\text{Forgacs, P., Horvath, Z., Palla, L.,} \textit{An 
exact fractionally charged self-dual solution}, Phys. Rev. Lett. 46 (1981), 
392. \\
 \hspace*{-0.75 in}\text{[FHP2]} \hspace*{.2 in}\text{Forgacs, P., Horvath, Z., Palla, L.,} \textit{One 
can have non-integer topological charge}, Z. Phys. C 12 (1982), 359-360.\\
 \hspace*{-0.75 in}\text{[FU] }  \hspace*{.3 in}\text{Freed, D.S., Uhlenbeck, K.,} \textit{Instantons and 
Four Manifolds}, Mathematical Sciences Research Institute Publications 1 (1984),
Springer-Verlag.\\
 \hspace*{-0.75 in}\text{[G]  }\hspace*{.48 in}\text{Giaquinta, M.,} \textit{Multiple Integrals in the Calculus
of Variations}, Annals of Mathematics Studies, Study 105 (1983).\\
 \hspace*{-0.75 in}\text{[H]  } \hspace*{.44 in}\text{Harland, D.,} \textit{Hyperbolic calorons, monopoles, and instantons}, Comm. Math. Phys. 280 (2008) 727-735.\\
 \hspace*{-0.75 in}\text{[JT] } \hspace*{.38 in}\text{Jaffe, A., Taubes, C.,} \textit{Vortices and Monopoles}, 
Birkhauser Progress in Physics 2 (1980).\\

\thispagestyle{fancy}
\newpage
\begin{center}
\vspace*{0.25 in}
\end{center}

\noindent\hspace*{-0.75 in}\text{[L]  } \hspace*{.46 in}\text{Landweber, G.D.,} \textit{Singular instantons with so(3) symmetry}, \\
arXiv:math.dg/0503611 (2005).\\
 \hspace*{-0.75 in}\text{[M]  } \hspace*{.38 in}Mazzeo, R., \textit{Unique continuation at infinity and embedded eigenvalues for asymptotically hyperbolic manifolds}, Amer. J. Math 113 (1991) 25-45.\\
 \hspace*{-0.75 in}\text{[MS] }\hspace*{.30 in} Manton, N., Sutcliffe, P., \textit{Topological Solitions}, Cambridge (2004).\\
   \hspace*{-0.75 in}\text{[PS] }\hspace*{.36 in} Prasad, M. K., Sommerfield, C. M., \textit{Exact classical solutions for the `t Hooft monopole and the Julia Zee dyon}, Phys. Rev. Lett. 35 (1975) 760-762.\\ 
 \hspace*{-0.75 in}\text{[R]  }\hspace*{.42 in} Rade, J., \textit{On singular Yang-Mills fields: 
gauge fixing and growth estimates}, preprint (1993).\\
 \hspace*{-0.75 in}\text{[RS] }\hspace*{.34 in} Reed, M., Simon, B., \textit{Analysis of Operators IV}, 
Academic Press (1978).\\
 \hspace*{-0.75 in}\text{[SS1]}\hspace*{.36 in} Sibner, L.M., Sibner, R.J., \textit{Singular Sobolev connections with holonomy}, Bull. Amer. Math. Soc. 19 (1988), 471-473.\\
 \hspace*{-0.75 in}\text{[SS2]}\hspace*{.36 in} Sibner, L.M., Sibner, R.J., \textit{Classification of 
singular Sobolev connections by their holonomy}, Comm. Math. Phys. 144 (1992), 
337-350.\\
 \hspace*{-0.75 in}\text{[SSU]}\hspace*{.34 in} Sibner, L.M., Sibner, R.J., Uhlenbeck, K., \textit{Solutions
to Yang-Mills Equations that are not self-dual}, Proc. Nat. Acad. Sci., USA 86 
(1989), 8610-8613.\\
 \hspace*{-0.75 in}\text{[T]  }\hspace*{.44 in} Taubes, C.H., \textit{Self-dual Yang-Mills connections 
on non-self-dual} 4\textit{-manifolds}, J. Diff. Geom. 17 (1982), 139-170.
\thispagestyle{plain}
\end{document}